\documentclass[english,aps,floats,onecolumn,showpacs,nofootinbib]{revtex4}
\usepackage{pslatex}
\usepackage[T1]{fontenc}
\usepackage[latin1]{inputenc}
\usepackage{graphicx}
\usepackage{epsfig}

\usepackage{calc}
\usepackage{ifthen}

{
{
{
\newcommand{\bea}{\begin{eqnarray}}
\newcommand{\eea}{\end{eqnarray}}

\newcommand{\nc}{\newcommand}
\nc{\renc}{\renewcommand}
\nc{\eqs}[2]{\mbox{Eqs.~(\ref{#1},\,\ref{#2})}}
\nc{\eq}[1]{\mbox{Eq.~(\ref{#1})}}
\nc{\figs}[2]{\mbox{Figs.~(\ref{#1},\,\ref{#2})}}
\nc{\fig}[1]{\mbox{Fig~.(\ref{#1})}}
\nc{\be}[1]{\begin{equation} \mbox{$\label{#1}$}}
\nc{\ee}{\vspace{0.1cm}\end{equation}}

\newcommand{\bean}{\begin{eqnarray*}}
\newcommand{\eean}{\end{eqnarray*}}

\def\GeV{{\rm \ GeV}}
\def\MeV{{\rm \ MeV}}

\def\TeV{{\rm \ TeV}}

\def\vphi{\varphi}

\def\lae{\;^{<}_{\sim} \;} \def\gae{\; ^{>}_{\sim} \;}

\begin{document}
\title{
New Q-ball Solutions in Gauge-Mediation, Affleck-Dine Baryogenesis and Gravitino Dark Matter}
\author{Francesca Doddato}
\email{f.doddato@lancaster.ac.uk}
\author{John McDonald}
\email{j.mcdonald@lancaster.ac.uk}
\affiliation{Consortium for Fundamental Physics, Cosmology and Astroparticle Physics Group, University of 
Lancaster, Lancaster LA1 4YB, UK}
%Revised 5/4/12
\begin{abstract}

      Affleck-Dine (AD) baryogenesis along a $d=6$ flat direction in gauge-mediated supersymmetry-breaking (GMSB) models can produce unstable Q-balls which naturally have field strength similar to the messenger scale. In this case a new kind of Q-ball is formed, intermediate between the gravity-mediated and gauge-mediated types. We study in detail these new Q-ball solutions, showing how their properties interpolate between standard gravity-mediated and gauge-mediated Q-balls as the AD field becomes larger than the messenger scale. It is shown that $E/Q$ for the Q-balls can be greater than the nucleon mass but less than the MSSM-LSP mass, leading to Q-ball decay primarily to Standard Model fermions. More significantly, if $E/Q$ is greater than the MSSM-LSP mass, decaying Q-balls can provide a natural source of non-thermal MSSM-LSPs, which can subsequently decay to gravitino dark matter without violating nucleosynthesis constraints. The model therefore provides a minimal scenario for baryogenesis and gravitino dark matter in the gauge-mediated MSSM, requiring no new fields.

\end{abstract}
\maketitle

\section{Introduction}

      Affleck-Dine baryogenesis is a particularly simple way to generate the baryon asymmetry in the MSSM \cite{ad}. However, in GMSB models, AD baryogenesis is severely constrained by Q-ball formation. For field strengths greater than the messenger mass, the potential is approximately flat. Q-balls forming in a flat potential have energy-per-charge $E/Q \propto Q^{-1/4}$, therefore for $Q$ large enough, $E/Q$ is less than the nucleon mass $m_{n}$. In this case, as a result of baryon number conservation, Q-balls cannot decay and would exist in the Universe at present \cite{ks}. Q-balls 
absorbed by neutron stars would destabilize the stars and so are observationally ruled out \cite{sw1,sw2}. Thus for AD baryogenesis to succeed in GMSB models, the Q-balls must be unstable. 

      In the MSSM with R-parity conservation, the lowest dimension flat directions which can support AD baryogenesis or leptogenesis are the $d = 4$ $(H_{u}L)^{2}$ and $d = 6$ $(u^{c}d^{c}d^{c})^{2}$ directions. $d = 4$ AD leptogenesis is a high reheating temperature ($T_R$) variant of AD baryogenesis, with $T_R \sim 10^8 \GeV$. In contrast, $d = 6$ AD baryogenesis is a low reheating temperature variant, with $T_R \lae 100 \GeV$, where the low reheating temperature is necessary to dilute the larger initial baryon density of the AD condensate. 

       $d = 4$ AD leptogenesis along the $(H_{u}L)^{2}$ direction is possible because the AD condensate decays and thermalizes at a temperature greater than that of the electroweak phase transition, $T_{EW}$. In this case the lepton asymmetry is transferred to a baryon asymmetry via $B + L$-violating sphaleron fluctuations\footnote{Even if Q-balls form along the $d = 4$ direction, they have a small enough charge to decay at $T > T_{EW}$.}. $d = 6$ AD baryogenesis is dynamically more complex. The AD condensate is unstable with respect to perturbations and fragments into condensate lumps, which subsequently evolve into Q-balls
 \cite{ks,km1,km2,km3,kk1,kk2,kk3,mitsuo}. The large field and baryonic charge of the $d = 6$ Q-balls protects them from thermalization and leads to a low decay temperature $T_{d} \lae 1 \GeV$ \cite{km1,km2}. As a result, $d = 6$ Q-balls can be a natural source of non-thermal NLSPs in the MSSM \cite{km2,km4}. The formation of large charge Q-balls also protects $d=6$ AD baryogenesis from the effect of additional lepton number violation in extensions of the MSSM, which could washout the baryon asymmetry from $d=4$ AD leptogenesis. More importantly, late-decaying Q-balls could provide a minimal scenario for gravitino dark matter in the MSSM, which requires a natural source of non-thermal NLSPs. 

   In this paper we will focus on the case of $d = 6$ AD baryogenesis. In \cite{fd1} we considered the case of AD baryogenesis along the $d = 6$ $(u^{c}d^{c}d^{c})^{2}$ of the MSSM in GMSB. We showed that if the gravitino mass is large enough, not much smaller than 1 GeV, then it is possible for the AD field at the onset of baryogenesis to be similar to the messenger scale when the condensate fragments. In this case, since the Q-balls are not forming on the logarithmic plateau of the GMSB flat direction potential, they can have $E/Q > m_{n}$ and so can decay before the era of nucleosynthesis, avoiding the problem of stable Q-balls. 

     In addition, late-decaying Q-balls open up new possibilities for 
gravitino dark matter in the gauge-mediated MSSM. Decay of thermal relic MSSM-LSPs\footnote{MSSM-LSP refers to the LSP of the MSSM sector i.e. excluding gravitinos and RH sneutrinos. The true LSP is assumed to be the gravitino. The MSSM-LSP is also the NLSP except in the case of a RH sneutrino NLSP.} as a source of gravitino dark matter appears to be generally ruled out by BBN constraints \cite{kaz}\footnote{Early studies of the effect of decaying particles on element abundances can be found in \cite{khlopov}.}. (See \cite{jasper} for a discussion of possible exceptions.) Therefore a non-thermal source of MSSM-LSPs is required, with the MSSM-LSPs being produced below their freeze-out temperature. 

   The BBN constraints on MSSM-LSP decay to gravitinos have been analyzed in detail in \cite{kaz}. The constraints follow from the decay of MSSM-LSPs at $T \lae T_{BBN} \approx 1 \MeV$, which can modify light element abundances formed during BBN. The results of \cite{kaz} can be summarized as follows. Non-thermal stau or sneutrino MSSM-LSPs can produce gravitino dark matter and remain consistent with BBN if the MSSM-LSP mass is greater than around 300 GeV and $m_{3/2} \lae 1 \GeV$ (see Figs.14 and 16 of \cite{kaz}), while a bino MSSM-LSP of mass greater than about 300 GeV is consistent with BBN if $m_{3/2} \lae 2 \times 10^{-2} \GeV$ (Figs. 9 and 10 of \cite{kaz}). There is a large hadronic component in the decay products of bino decay, which leads to stronger constraints than the case of a stau or sneutrino MSSM-LSP, which have primarily 
radiative decays with only a small hadronic component. (The stau can also catalyze formation of $\;^{6}$Li via formation of $^{4}$He-$\tilde{\tau}$ bound states.) The analysis of \cite{kaz} considers only two MSSM-LSP masses, 100 GeV and 300 GeV. In general, there is no simple relation between the gravitino mass upper bound from BBN and the MSSM-LSP mass. The upper bound generally becomes weaker for larger MSSM-LSP mass. This can be understood as being due to the earlier time of decay. We can roughly estimate how the BBN bounds will depend on $m_{{\rm MSSM-LSP}}$ and $m_{3/2}$ by assuming that the same total injected energy at a given decay time will have a similar effect on element abundances. The MSSM-LSP decay rate depends on the MSSM-LSP mass and gravitino mass as $\Gamma_{{\rm MSSM-LSP}} \propto m_{{\rm MSSM-LSP}}^{5}/m_{3/2}^2$. We therefore expect the BBN gravitino mass bounds to vary as $m_{{\rm MSSM-LSP}}^{5/2}$ for a fixed value of the product of the MSSM-LSP mass and its abundance. This roughly fits the trend in the figures in \cite{kaz}. However, closer inspection shows that such a simple scaling does not work exactly, as the bound from each light element has a different dependence on the MSSM-LSP mass. 

    Since $d = 6$ Q-balls typically decay below the MSSM-LSP freeze-out temperature, they can provide a natural source of non-thermal MSSM-LSPs in the gauge-mediated MSSM. Alternatively, if RH sneutrino NLSPs have the right properties, Q-ball decay to RH sneutrinos might produce gravitino dark matter while remaining consistent with BBN even if $m_{3/2} > 1 \GeV$. 

    In the case where only positive charged Q-balls result from condensate fragmentation, B-conservation combined with R-parity conservation implies that\footnote{In the context of gravity-mediated SUSY breaking, Q-ball decay to 2 GeV axino LSPs was proposed in \cite{rs} in order to satisfy the required LSP mass.} $m_{3/2} \approx 2 \GeV$. This can be seen as follows. 
Each unit of $Q$ corresponds to an R-parity odd squark absorbed by the Q-ball. Therefore $|\Delta Q| = 1$ corresponds to a change of sign of the R-parity of the Q-ball, and so a MSSM-LSP must be emitted by the Q-ball. Thus B-conservation combined with R-parity conservation implies that $n_{\chi} = 3 n_{B}$ from Q-ball decay, where $\chi$ denotes the MSSM-LSP.  (In the following the global charge $Q$ of the Q-ball squarks is normalized to 1, so that $B = Q/3$.) Subsequently, each MSSM-LSP decays to a gravitino. The gravitino mass density is therefore related to the baryon mass density by $\rho_{3/2} = m_{3/2} n_{\chi} = 3 (m_{3/2}/m_{n}) \rho_{B}$, where $m_{n}$ is the nucleon mass. Therefore $m_{3/2} = (m_{n}/3)(\Omega_{3/2}/\Omega_{B})$. Assuming gravitinos account for the observed cold dark matter abundance, $\Omega_{3/2}/\Omega_{B} \approx 5$, this then implies that $m_{3/2} \approx 2 \GeV$. This value is self-consistent with the large gravitino mass required to have unstable Q-balls.  

         A successful RH sneutrino NLSP scenario with $m_{3/2} \approx 2 \GeV$ requires that the MSSM-LSPs from Q-ball decay can decay to RH sneutrinos before nucleosynthesis and that the RH sneutrinos can decay to gravitinos without violating nucleosynthesis and free-streaming constraints \cite{rhsn}, which might be achieved if the RH neutrinos have enhanced Yukawa couplings via the see-saw mechanism \cite{fdc}.  

     Alternatively, smaller dark matter gravitino masses might be possible if both positively and negatively charged Q-balls can form. As we will discuss, this is natural in GMSB and can relieve the tension between the dark matter gravitino mass and the BBN upper bound. 

    The analysis of Q-ball decay in \cite{fd1} was based on the assumption that the GMSB Q-balls with $\varphi(0) \sim M_m$, where $\varphi(0)$ is the field strength at the centre of the Q-ball, could be approximated by gravity-mediated-type Q-balls. However, Q-balls in the transition region between the $|\Phi|^2$ potential at $|\Phi|/M_{m} \ll 1$ and the approximately constant (logarithmic) potential at $|\Phi|/M_{m} \gg 1$ will be of a new type\footnote{We should distinguish these Q-balls from another Q-ball solution which interpolates between gauge and gravity-mediated Q-balls  \cite{kkgm}. These were derived for the case of large AD field, where the gravity-mediated contribution 
to the potential comes to dominate the gauge-mediated contribution.}, interpolating between gravity-mediated type with $E/Q$ approximately constant and gauge-mediated type with $E/Q \propto Q^{-1/4}$. Since the value of $E/Q$ determines of the decay properties of the Q-balls, in particular their decay temperature and whether MSSM-LSPs are produced in Q-ball decay, it is important to understand these Q-balls in detail. In this paper we will study the Q-ball solutions in the transition region as a function of $\varphi(0)/M_m$.

    The paper is organized as follows.  In Section 2 we discuss the flat-direction potential which models a generic GMSB flat-direction. In Section 3 we derive the equation for the Q-ball solutions. In Section 4 we show that the Q-ball solutions have a useful scaling property, broken only by small gravity-mediated contributions. In Section 5 we present our results for the properties of the Q-balls as a function of $\varphi(0)/M_m$. In Section 6 we discuss the implications of our results for AD baryogenesis and gravitino dark matter. In Section 7 we present out conclusions.

\section{Flat-direction potential in GMSB}

   GMSB models are based on SUSY breaking in a hidden sector which is transmitted to the MSSM via vector pairs of messenger fields carrying SM gauge charges. The messenger superfield scalar components acquire SUSY breaking mass splittings from their interaction with the hidden sector. The messengers then induce masses for MSSM gauginos at 1-loop and soft SUSY breaking scalar mass squared terms at 2-loops. A key relation is that between the gravitino mass and the messenger mass, given by \cite{fd1}
\be{e0} M_{m} \approx \frac{g^2}{16 \pi^2} \frac{\sqrt{3} \kappa m_{3/2} M_{p}}{m_{s}}    ~,\ee
where $g$ is the gauge coupling of the messengers, $\kappa$ is the superpotential coupling of the messengers to the SUSY breaking field, $m_{s}$ is the soft SUSY breaking scalar mass and $M_{p} =  2.4 \times 10^{18} \GeV$ is the reduced Planck mass. 
Therefore, assuming minimal SUGRA,  
\be{e0a}  M_{m} \approx 5 \times 10^{13} g^2 \kappa \left( \frac{m_{3/2}}{2 \GeV} \right) 
\left( \frac{100 \GeV}{m_{s}} \right)  \GeV      ~.\ee
Therefore large gravitino masses are required for large $M_{m}$.   

   Q-balls form in a scalar field theory with a global $U(1)$ symmetry, when the potential $V(|\Phi|)$ is flatter than  
$|\Phi|^2$, corresponding to an attractive interaction between the scalar particles. In the case of GMSB models, the form of the flat direction potential in the region of $|\Phi|$ of interest is \cite{deg,fd1}
\be{e1}  V(\Phi) = m_{s}^2 M_{m}^2 \ln^{2} \left(1 + \frac{|\Phi|}{M_{m}}\right)\left(1 + K \ln \left( \frac{|\Phi|^2}{M_{m}^2} \right) \right) + m_{3/2}^2\left(1 + \hat{K} \ln \left( \frac{|\Phi|^2}{M_{m}^2} \right) \right)|\Phi|^2 ~.\ee 
In this we assume $|\Phi|$ is small enough that the $U(1)$-violating A-terms and the non-renormalizable potential terms of the full GMSB potential can be neglected. The former condition is essential for the existence of Q-balls.  
The first term in \eq{e1} is due to GMSB with messenger mass $M_{m}$. The factor multiplying this takes into account 1-loop radiative corrections due to gaugino loops once $|\Phi| \lae M_{m}$, with $K \approx -(0.01-0.1)$ \cite{km1,km2,km3}. The second term is due to gravity-mediated SUSY breaking including the 1-loop correction term $ \hat{K}$. (For simplicity we set $\hat{K} = K$.) In the limit $|\Phi|/M_{m} \ll 1$, the potential 
\eq{e1} tends towards a gravity-mediated-type potential of the form 
\be{e3} V(|\Phi|) \approx m_{s}^{2} \left(1 + K \ln \left( \frac{|\Phi|^2}{M_{m}^2} \right) \right) |\Phi|^2 + 
   m_{3/2}^2\left(1 + \hat{K} \ln \left( \frac{|\Phi|^2}{M_{m}^2} \right) \right)|\Phi|^2    ~,\ee
while in the limit $|\Phi|/M_{m} \gg 1$ the potential has a slow logarithmic growth with $|\Phi|$ (the GMSB 'plateau') plus a small contribution from gravity-mediated SUSY breaking, 
\be{e4} V(|\Phi|) \approx m_{s}^{2} M_{m}^{2} \ln^{2}\left(\frac{|\Phi|}{M_{m}}\right) + 
   m_{3/2}^2\left(1 + \hat{K} \ln \left( \frac{|\Phi|^2}{M_{m}^2} \right) \right)|\Phi|^2   ~.\ee
(The term proportional to $K$ is a small correction to the potential in this case and may be neglected.) Q-ball solutions at $\varphi(0)/M_{m} \ll 1$ will have the form of gravity-mediated Q-balls, with a Gaussian profile $\varphi(r)$ and constant $E/Q \approx m_{s}$. At $|\Phi|/M_{m} \gg 1$, the potential is not exactly constant but the Q-balls are expected to be similar to those for a perfectly constant flat direction potential, which have
$E/Q \propto Q^{-1/4}$ and therefore have a suppressed $E/Q$ value at large enough $Q$. 

  In the case of $d = 6$ flat directions in GMSB, when $m_{3/2} = 2 \GeV$, which is consistent with gravitino dark matter from Q-ball decay in the case where only positively charged Q-balls form, the onset of oscillations of the $\Phi$ field occurs when 
$|\Phi|$ is close to or somewhat larger than the messenger mass. 
The precise value at which oscillations begin is sensitive to the non-renormalizable B-violating superpotential operator which lifts the flat direction. This is assumed to be of the form $W = \Phi^6/6! \tilde{M}^3$, where $\tilde{M}$ is assumed close to the Planck scale, $\tilde{M} = (0.1-1)M_{p}$.  
There is then a period of time during which quantum fluctuations of the AD field grow and $|\Phi|$ decreases due to the expansion of the Universe, eventually breaking up the condensate into fragments carrying large baryon number. As a result, for typical parameters in \eq{e1}, the value of $|\Phi|/M_{m}$ in the fragments is typically in the transition region between $|\Phi|/M_{m} \ll 1$ and $|\Phi|/M_{m} \gg 1$. These fragments subsequently evolve into Q-balls. The evolution of the fragments into Q-balls is determined by the non-linear dynamics of $\Phi$, and may 
result in the production of purely positive or both positive and negative charged Q-balls \cite{qbf}.  

  The value of $E/Q$ is crucial for determining the decay properties of the Q-balls. Each unit of $Q$ corresponds to an R-parity odd squark absorbed by the Q-ball. Therefore $|\Delta Q| = 1$ corresponds to a change of sign of the R-parity of the Q-ball, and so a MSSM-LSP must be emitted by the Q-ball. To have Q-ball decay to MSSM-LSPs which is not kinematically suppressed, we must therefore have $E/Q > m_{\chi}$, otherwise the decay would have to occur through $|\Delta Q| = 2$ processes to R-parity even final states of SM quarks and leptons. This process may be thought of as squarks in the Q-ball annihilating with each other to quark pairs, rather than individually decaying to a quark plus MSSM-LSP.  This is a new mode of Q-ball decay which allows a new scenario for AD baryogenesis in GMSB models, with Q-balls decaying at low temperature to baryon number without accompanying MSSM-LSPs. This possibility has also been noted in \cite{kknew}, where it plays a central role in an alternative model for gravitino dark matter from Q-ball decay.    

       The ratio of Q-ball decay via the $|\Delta Q| = 2$ annihilation process relative to the conventional $|\Delta Q| = 1$ decay process can be estimated by comparing the decay rate of the $\Phi$ scalars inside the Q-ball to the annihilation rate of $\Phi$ pairs, using $\Gamma_{ann} \approx n  \sigma v$, where $n = \omega \varphi(0)^2$ is the charge density in the Q-ball and $\sigma v$ is the non-relativistic annihilation cross-section times relative velocity \cite{kknew}.  For the case of $\Phi$ decay to quark plus gluino and $\Phi$ annihilation to quark pairs via gluino exchange, we find $\Gamma_{decay} \approx 2 \Gamma_{ann}$. This can increase the dark matter gravitino mass for positive Q-balls by an O(1) factor, since the number of MSSM-LSPs and so gravitinos produced in Q-ball decay can be reduced, requiring a larger gravitino mass to account for dark matter.

\section{Q-ball solutions for GMSB Flat Directions}

    Q-balls are minimum energy configurations for a fixed global charge $Q$. The solutions are obtained by introducing a Lagrange multiplier $\omega$ for the conserved charge and minimizing the functional 
\be{e5} E_\omega(\Phi, \dot{\Phi}, \omega) = E + \omega\left(Q - \int d^3x\;\rho_Q \right)  ~,\ee
where 
\be{e5a}   E =  \int d^{3}x \; |\dot{\Phi}|^2 + |\nabla \Phi|^2 +V(|\Phi|)    ~\ee
and 
\be{e5b}  \rho_{Q} = i \left(\dot{\Phi}^{\dagger} \Phi
 - \dot{\Phi}^{\dagger}\Phi \right)         ~\ee
The minimum energy solutions have the form 
\be{e6} \Phi = \frac{\varphi(r)}{\sqrt{2}}e^{i \omega t} ~,\ee
where the Q-ball profile $\vphi(r)$ is given by the solution of 
\be{e7}  \frac{\partial^2 \varphi}{\partial r^2} + \frac{2}{r}\frac{\partial \varphi}{\partial r} = \frac{\partial V}{\partial \varphi} - \omega^2 \varphi  ~,\ee
with boundary conditions $\vphi^{'}(r) = 0 $ as $r \rightarrow 0$ and $\vphi(r) \rightarrow 0$ as $r \rightarrow \infty$.

    For the GMSB potential \eq{e1}, the equation for the Q-ball profile is 
\be{e7a}  \frac{\partial^{2} \vphi}{\partial r^2} + \frac{2}{r} \frac{\partial \vphi}{\partial r} = 
m_s^2 M_{m}^2 \frac{ \sqrt{2} \ln \left(1 + \frac{\vphi}{\sqrt{2} M_m}\right) }{\left(1 + \frac{\vphi}{\sqrt{2} M_m}\right) } \left( 1 + K \ln\left(\frac{\vphi}{2 M_m} \right) \right) 
+ \frac{2 K}{\vphi} \ln^{2}\left(1 + \frac{\vphi}{\sqrt{2} M_m} \right) + m_{3/2}^2 \vphi 
 \left( 1 + K \ln \left(\frac{\vphi}{2 M_m} \right) \right) - \omega^2 \vphi    ~.\ee  
Our method to solve \eq{e7a} is to fix the value of $\vphi(0)/M_m$ and integrate \eq{e7a} with the boundary condition $\vphi^{'}(0) = 0$. We then vary $\omega$ until $\vphi(r) \rightarrow 0$ as $r \rightarrow \infty$. This determines the Q-ball profile $\varphi(r)$ and the value of $\omega$ for the given value of $\vphi(0)/M_{m}$.

The total charge of the Q-ball is then 
\be{e8} Q = \int^{\infty}_{0} 4\pi r^2 \omega \varphi(r)^2 dr  ~,\ee
while the total energy is 
\be{e9}  E = \int^{\infty}_{0}  4 \pi r^2 \left[ \frac{1}{2} \left(\frac{\partial \varphi}{\partial r}\right)^2 
+ V(\varphi) + \frac{\omega^2 \varphi^2}{2} \right] dr. ~\ee 
Using these we can compute $E/Q$ for the Q-ball.  The radius of the Q-ball presented in our tables is defined to be the radius within which 90$\%$ of the total energy is found.
     
        In the limit 
$\vphi(0)/M_m \ll 1$, the Q-balls are of gravity-mediated type, in which case an exact solution exists. For the potential 
\be{e10}   V(\Phi) =  m_{s}^{2}  \left(1 + K \ln \left( \frac{|\Phi|^2}{M_{m}^2} \right) \right) |\Phi|^2      ~,\ee
the solution for $\vphi(r)$ is  
\be{e11}  \vphi(r) = \vphi(0)e^{-r^{2}/R^{2}}      ~\ee
where 
$$ R^{2} = \frac{2}{|K|m_{s}^2} \;\;\; ; \;\;\; \omega^2 = \omega^2_{o} + m_{s}^{2}(1 + K) \;\;;\;\; \omega_{o}^{2} = 3 |K| m_{s}^{2} + K \ln \left( \frac{\vphi(0)^2}{M_{m}^2} \right)    ~.$$
Then $E/Q \approx \omega \approx m_{s}$ when $|K| \ll 1$, independent of $\vphi(0)/M_m$. In the opposite limit, $\vphi(0)/M_m \gg 1$, if the potential is treated as constant then $E/Q \propto Q^{-1/4}$.

    A key quantity for AD baryogenesis is the Q-ball decay temperature. Q-ball decay to fermions has an upper bound from 
Pauli blocking \cite{qbd}. This will generally be saturated since $\varphi(0) \gg \omega$ \cite{qbd}, therefore 
\be{e12} T_{d} =  \left( \frac{\omega^{3} R^{2} M_{Pl}}{48 \pi k_{T} Q} \right)^{1/2}     ~,\ee
where $k_{T} = (g(T)\pi^{2}/90)^{1/2}$ and $g(T)$ is the effective number of relativistic degrees of freedom, with the expansion rate during radiation domination given by $H = k_{T} T^2 /M_{p}$.

\section{Scaling properties of the Q-ball solutions}

  The solutions for any $m_{s}$ and $M_{m}$ can be obtained by rescaling the solution of \eq{e7a}, up to small corrections 
from the gravitino mass, which breaks scale-invariance. This can be seen by expressing the Q-ball equation in terms of 
$\hat{\vphi} = \vphi/M_m$, $\hat{r} = r m_{s}$ and $\hat{\omega} = \omega/m_s$. 
\eq{e7a} then becomes 
\be{e13}  \frac{\partial^{2} \hat{\vphi}}{\partial \hat{r}^2} + \frac{2}{\hat{r}} \frac{\partial \hat{\vphi}}{\partial \hat{r}} = 
\frac{ \sqrt{2} \ln\left(1 + \frac{\hat{\vphi}}{\sqrt{2}}\right) }{\left(1 + \frac{\hat{\vphi}}{\sqrt{2}}\right) } \left( 1 + K \ln \left(\frac{\hat{\vphi}}{2} \right) \right) 
+ \frac{2 K}{\hat{\vphi}} \ln^{2}\left(1 + \frac{\hat{\vphi}}{\sqrt{2}} \right) + \left(\frac{m_{3/2}}{m_{s}}\right)^2 \hat{\vphi} 
 \left( 1 + K \ln\left(\frac{\hat{\vphi}}{2} \right) \right) - \hat{\omega}^2 \hat{\vphi}    ~.\ee  
In the limit $m_{3/2} \rightarrow 0$ this has a unique solution for a given $K$ and $\varphi(0)/M_m$. For a given $m_s$ and $M_m$ this can then be transformed to the physical solution using $\varphi(r) = M_{m}\hat{\varphi}(m_{s} r)$. The energy and charge of the Q-ball scale as $E \propto M_{m}^{2}/m_{s}$ and $Q \propto M_{m}^{2}/m_{s}^{2}$, as can be seen by expressing \eq{e8} and \eq{e9} in terms of $\hat{\varphi}$, $\hat{r}$ and $\hat{\omega}$. Therefore $E/Q \propto m_{s}$ and is independent of $M_m$. These properties are confirmed by the results in Tables 1-3. Using these scaling properties and the results in the Tables, we can obtain accurate estimates of the Q-ball properties for any $m_{s}$ and $M_{m}$, up to small corrections of order $\left(m_{3/2}/m_{s}\right)^2$. In particular, applying this scaling to the Q-ball decay temperature gives $T_d \propto m_{s}^{3/2}/M_m$.

\begin{figure}[htbp]
\begin{center}
\epsfig{file=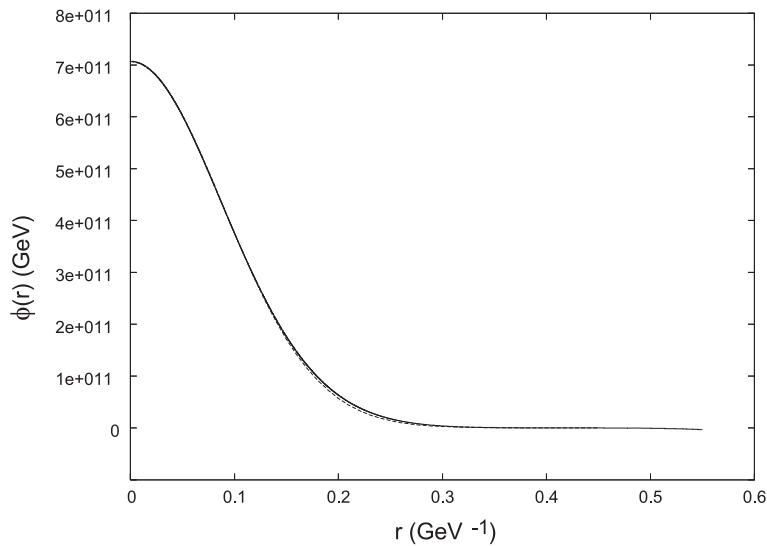, width=0.4\textwidth, angle = -90}
\caption{Q-Ball Solution for $\varphi(0)/M_m = 0.01$, $K = -0.01$, $m_{3/2} = 2 \GeV$, $M_m = 10^{14} \GeV$ and  $m_s = 100 \GeV$. The corresponding gravity-mediated Q-ball is shown as a dashed line.}
\label{fig1}
\end{center}
\end{figure}

\begin{figure}[htbp]
\begin{center}
\epsfig{file=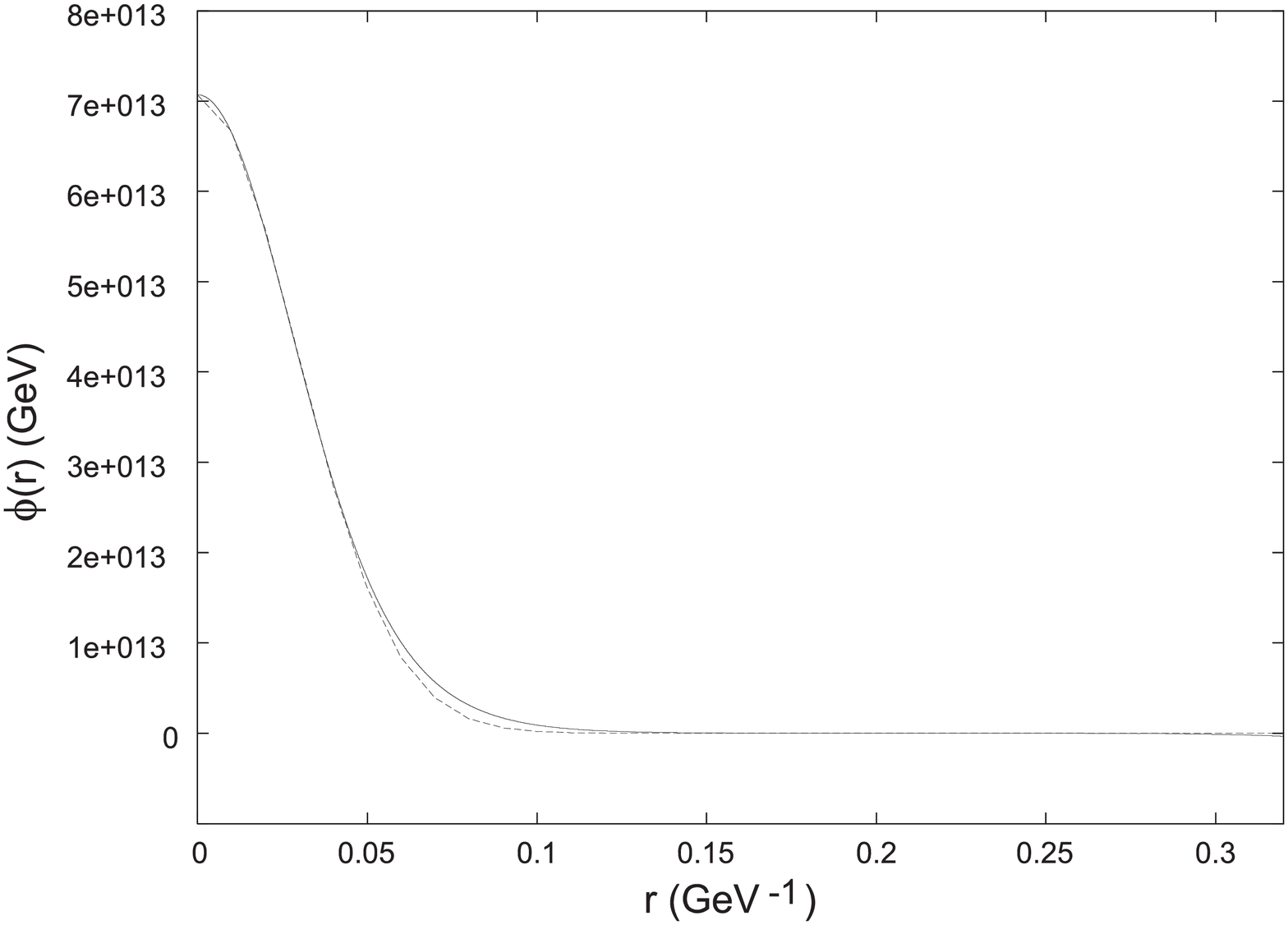, width=0.4\textwidth, angle = -90}
\caption{Q-Ball Solution for $\varphi(0)/M_m = 1$, $K = -0.01$, $m_{3/2} = 2 \GeV$, $M_m = 10^{14} \GeV$ and $m_s = 100 \GeV$. The corresponding gravity-mediated Q-ball is shown as a dashed line.}
\label{fig2}
\end{center}
\end{figure}

\begin{figure}[htbp]
\begin{center}
\epsfig{file=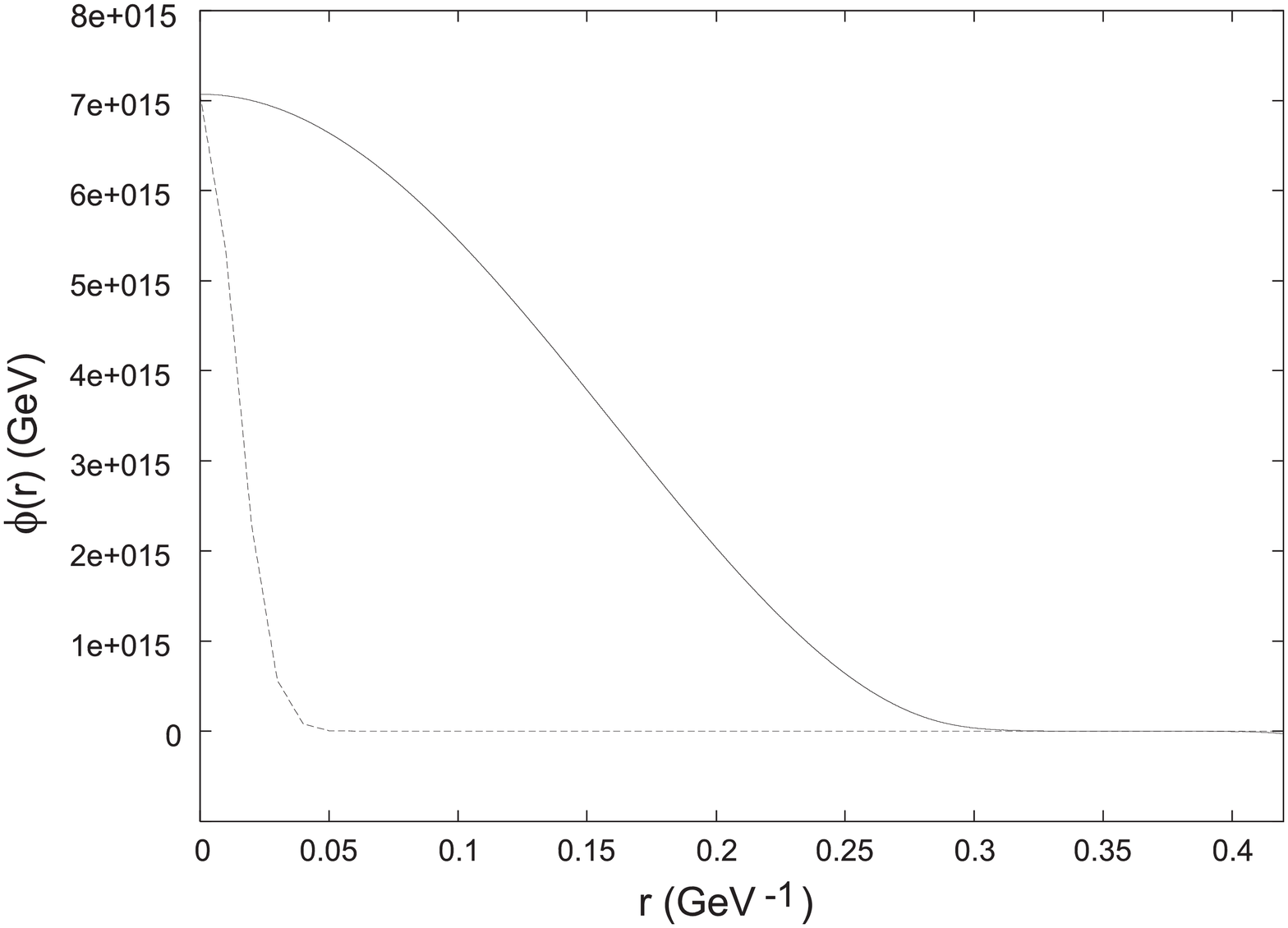, width=0.4\textwidth, angle = -90}
\caption{Q-Ball Solution for $\varphi(0)/M_m = 100$, $K = -0.01$, $m_{3/2} = 2 \GeV$, $M_m = 10^{14} \GeV$ and $m_s = 100 \GeV$. The corresponding gravity-mediated Q-ball is shown as a dashed line.}
\label{fig3}
\end{center}
\end{figure}

\begin{figure}[htbp]
\begin{center}
\epsfig{file=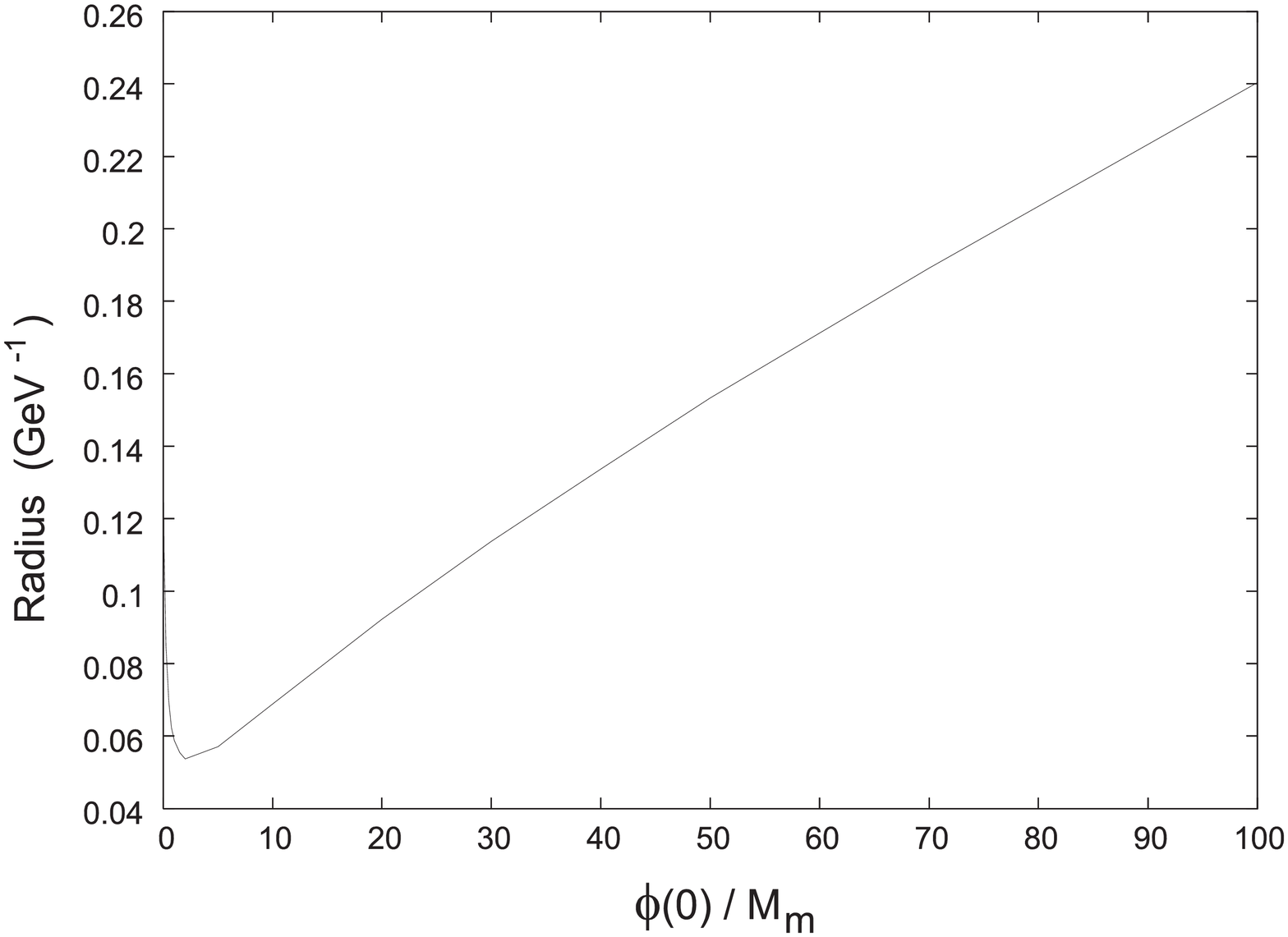, width=0.4\textwidth, angle = -90}
\caption{The variation of the radius of the Q-ball for $K = -0.01$, $m_{3/2} = 2 \GeV$, $M_m = 10^{14} \GeV$ and $m_s = 100$ $\GeV$.}
\label{fig4}
\end{center}
\end{figure}

\begin{figure}[htbp]
\begin{center}
\epsfig{file=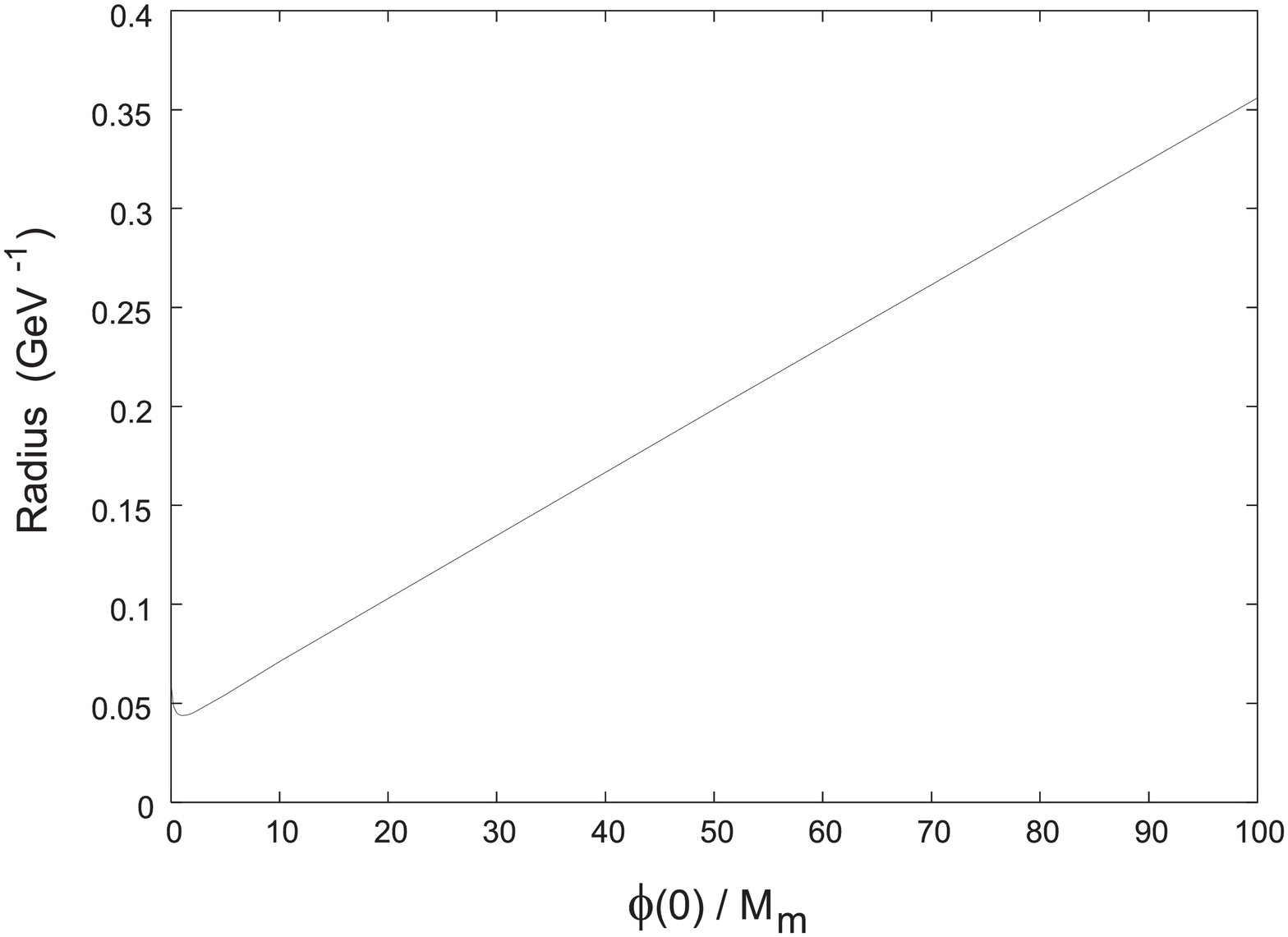, width=0.4\textwidth, angle = -90}
\caption{The variation of the radius of the Q-ball for $K = -0.1$, $m_{3/2} = 2 \GeV$, $M_m = 10^{14} \GeV$ and $m_s = 100$ $\GeV$.}
\label{fig5}
\end{center}
\end{figure}

\begin{figure}[htbp]
\begin{center}
\epsfig{file=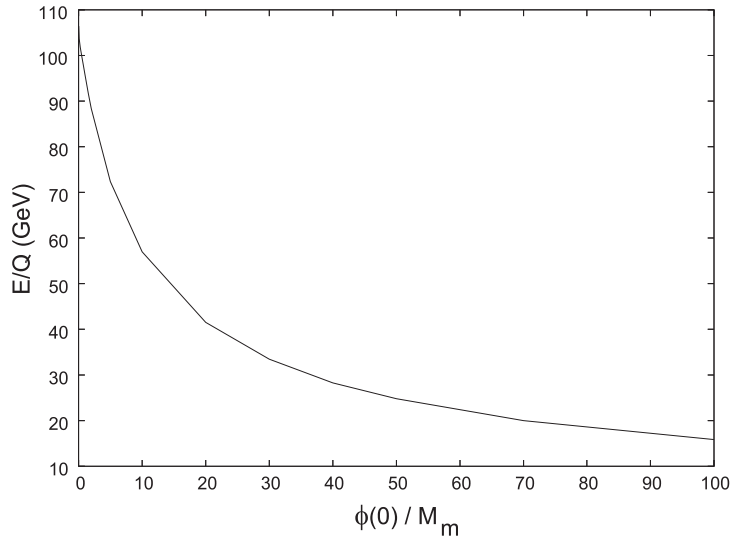, width=0.4\textwidth, angle = -90}
\caption{The variation of E/Q as $\varphi(0)/M_m$ grows, with $K = -0.01$, $m_{3/2} = 2 \GeV$, $M_m = 10^{14} \GeV$ and  $m_s = 100 \GeV$.}
\label{fig6}
\end{center}
\end{figure}

\begin{figure}[htbp]
\begin{center}
\epsfig{file=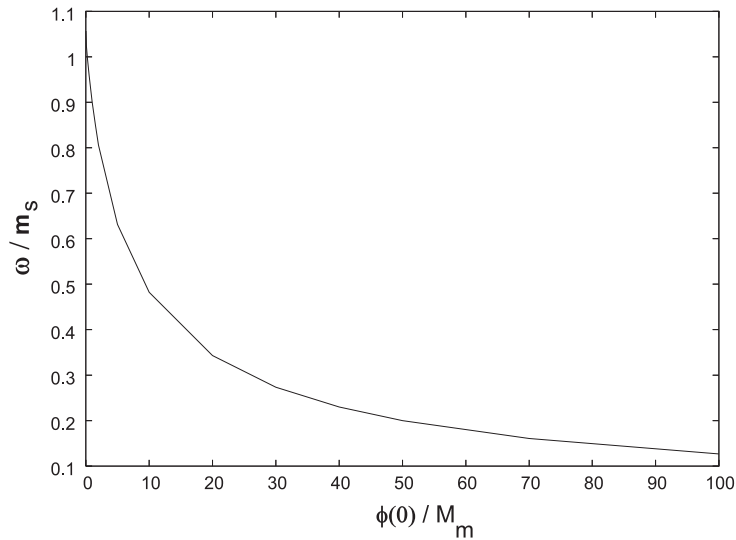, width=0.4\textwidth, angle = -90}
\caption{$\omega/m_{s}$ vs. $\varphi(0)/M_m$ for $K = -0.01$, $M_m = 10^{14} \GeV$, $m_{3/2} = 2 \GeV$ and $m_s = 100$ $\GeV$.}
\label{fig7}
\end{center}
\end{figure}

\begin{figure}[htbp]
\begin{center}
\epsfig{file=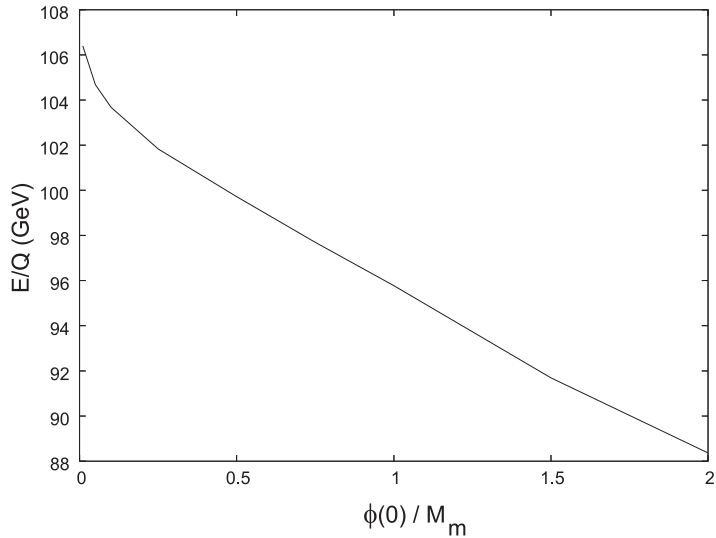, width=0.4\textwidth, angle = -90}
\caption{The variation of E/Q as $\varphi(0)/M_m$ grows, close-up on the transition region, $\varphi(0)/M_m = 0.01$ to $\varphi/M_m = 2$.}
\label{fig8}
\end{center}
\end{figure}

\begin{figure}[htbp]
\begin{center}
\epsfig{file=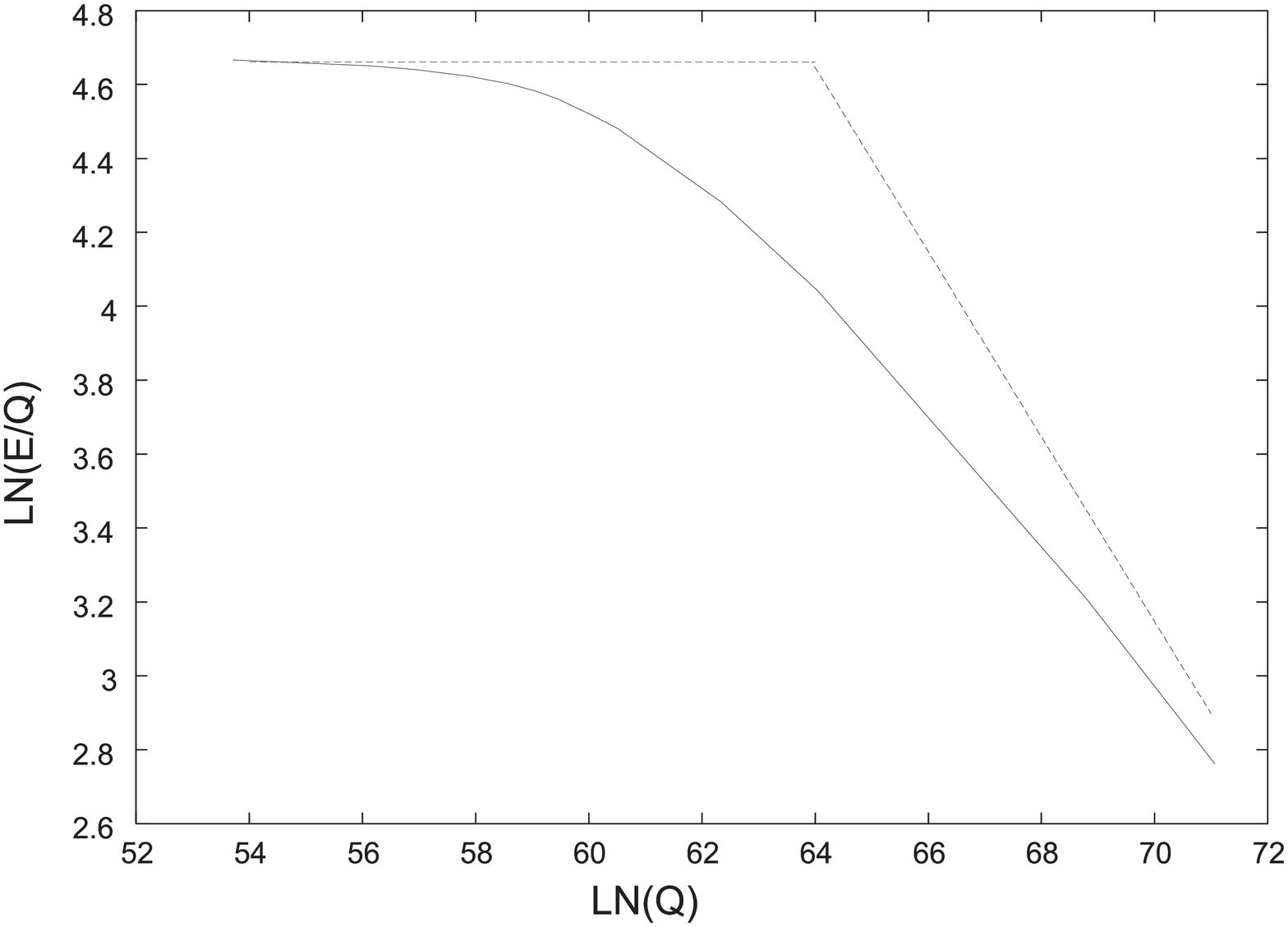, width=0.4\textwidth, angle = -90}
\caption{LN(E/Q) vs. LN(Q) for $K = -0.01$, $M_m = 10^{14} \GeV$, $m_{3/2} = 2 \GeV$ and $m_s = 100 \GeV$. The constant potential approximation is shown as a dashed line.}
\label{fig9}
\end{center}
\end{figure}

\section{Results}

      In Figures 1-3 we show the Q-ball profile $\varphi(r)/M_m$ for the cases $\varphi(0)/M_{m} = 0.1$, 1 and 100, when $K = -0.01$  and $m_{s} = 100 \GeV$. (We fix $M_m = 10^{14} \GeV$ and $m_{3/2} = 2 \GeV$ throughout.)  The gravity-mediated Q-ball is shown as a dashed line for comparison. For $\varphi(0)/M_m = 0.1$ and 1, the Q-ball closely matches the Gaussian 
profile expected for a gravity-mediated-type Q-ball. For $\varphi(0)/M_m \gg 1$ the profile deviates strongly from Gaussian, with a much wider profile for a given $\varphi(0)/M_m$. 

   In Figures 4 and 5 we show the Q-ball radius as a function of $\varphi(0)/M_m$ for $K = -0.01$ and $K = -0.1$. This shows the Q-ball radius decreasing slightly as $\varphi(0)/M_m$ approaches 1, then increasing almost linearly with $\varphi(0)/M_m$ as it becomes larger than 1. 

      In Figure 6 we show $E/Q$ as a function of $\varphi(0)/M_m$ for $K = -0.01$ and $m_s = 100 \GeV$. (The corresponding value of $\omega$ for the Q-ball solutions is shown in Figure 7.) A close-up of $E/Q$ as a function of $\varphi(0)/M_m$ around $\varphi(0)/M_m \sim 1$ is shown in Figure 8. $E/Q$ decreases from $E/Q \approx m_s$ to $E/Q \approx 0.15 m_s$ as $\varphi(0)/M_m$ increases from 0.01 to 100. Therefore, even for large $\varphi(0)/M_m$, $E/Q$ can be larger than the nucleon mass and Q-balls can be unstable. Even larger values of $\varphi(0)/M_m$ would be allowed with larger $m_s$. Thus $d=6$ AD baryogenesis can easily produce unstable Q-balls, even though $E/Q$ can be significantly suppressed relative to $m_s$. 

In Figure 9 we show $E/Q$ as a function of $Q$. We also show the approximate solution assuming a purely gravity-mediated potential when $\varphi(0)/M_m < 1$ and a constant potential once $\varphi(0)/M_m > 1$. For $\ln(Q) \lae 60$, corresponding to $\varphi(0)/M_m \lae 1$, $E/Q$ becomes independent of $Q$, as expected for a gravity-mediated-type Q-ball. At larger $Q$ we find an almost constant negative slope, corresponding to $E/Q \propto Q^{-n}$ with $n = 0.22$. This differs slightly from the conventional value for gauge-mediated-type Q-balls, $n = 0.25$. However, $n = 0.25$ assumes a completely flat potential, whereas in our case there is a squared logarithmic potential. 

    In Tables 1-3 we give the numerical properties of the Q-ball solutions as a function of $\varphi(0)/M_m$ for the cases $K = -0.01$, $m_s = 100 \GeV$ (Table 1), $K = -0.01$, $m_s = 1 \TeV$ (Table 2) and  $K = -0.1$, $m_s = 100 \GeV$ (Table 3). In particular, we show $Q$ and $T_d$ as a function of $\varphi(0)/M_m$. The tables assume $M_{m} = 10^{14} \GeV$; the results can be scaled to other values of $m_{s}$ and $M_{m}$ by using the scaling properties discussed in the previous section. We see that a large baryonic charge is typical for the Q-balls, with $Q \gae 10^{21}$ for the examples given in the tables. As $\varphi(0)/M_m$ increases, the charge rapidly increases. As a result $T_{d} \propto Q^{-1/2}$ rapidly decreases. 
For the case $M_m = 10^{14} \GeV$ and $m_s = 100 \GeV$ in Table 1, $T_d$ becomes less than the nucleosynthesis bound $\sim$ 1 MeV once $\varphi(0)/M_m \gae 1$. Smaller values of $M_m$ and larger values of $m_s$
allow larger $\varphi(0)/M_m$ to be compatible with $T_d \gae 1 \MeV$, since $T_d$ scales as $T_d \propto m_{s}^{3/2}/M_m$. So $M_m = 10^{13} \GeV$ and $m_s = 500 \GeV$ enhances $T_d$ by a factor 110, in which case $\varphi(0)/M_m$ can be as large as 20. However, in general there is an upper limit on $\varphi(0)/M_m$ from the Q-ball decay temperature, which we expect 
to be no larger than O(100) in plausible models.  This is an important constraint, as it means that the value of $|\Phi|$ at the onset of AD baryogenesis cannot be arbitrarily large compared to the messenger scale. Nevertheless, it is possible for  $|\Phi|$ to be significantly larger than the messenger scale, which can allow for smaller messenger scales and so gravitino masses. It also means that in many cases $\varphi(0)$ will not be very much larger than the messenger scale. Comparing Tables 1 and 3, we see that increasing $|K|$ from 0.01 to 0.1 also has a small effect on $T_d$, increasing $T_d$ by a factor of 2 at $\varphi(0)/M_m \lae 1$, less so at larger $\varphi(0)/M_m$.

\begin{table}[h]
\begin{center}
\begin{tabular}{|c|c|c|c|c|c|c|c|c|}
\hline $\varphi(0) / M_m$ & $\omega / m_s$ & $E$ (GeV) & $Q$ & $E/Q$ (GeV) & $ln(E/Q)$ & $ln(Q)$	&	$Radius$ $({\rm GeV^{-1}})$	&	$T_d$ (GeV)\\
\hline $0.01$  & $1.056834897$	&	$2.2440E+25$ & $2.109E+23$ & $1.0640E+02$  &	$4.667216272$ &	$53.70567104$ &	$0.1622$ &	$0.047177955$	\\
\hline $0.05$  & $1.036387958$	& $2.4590E+26$ & $2.349E+24$ & $1.0468E+02$	& $4.650935244$	& $56.11603194$ & $0.1303$ &	$0.01102814$	\\
\hline $0.1$	 & $1.023327904$	& $5.4890E+26$ & $5.294E+24$ & $1.0368E+02$	& $4.641342172$	& $56.92861634$	& $0.1107$ &	$0.006123415$	\\
\hline $0.25$	 & $0.99574093$	& $1.3880E+27$ & $1.363E+25$ & $1.0183E+02$	& $4.623345895$ & $57.87431548$	& $0.0849$ &	$0.002809281$	\\
\hline $0.5$   & $0.959374796$	& $2.8080E+27$ & $2.816E+25$ & $9.9716E+01$  &	$4.602325234$	& $58.59994476$	& $0.06937$ &	$0.001510267$	\\
\hline $0.75$	 & $0.92784697$	& $4.4120E+27$ & $4.516E+25$ & $9.7697E+01$	& $4.581871641$	& $59.07225397$ &	$0.06239$ &	$0.00102016$	\\
\hline $1.0$   & $0.899333086$	& $6.2800E+27$ & $6.557E+25$ & $9.5776E+01$  &	$4.562006985$ &	$59.44516051$ & $0.05874$ &	$0.000760638$	\\
\hline $1.5$   & $0.849117189$	& $1.0920E+28$ & $1.191E+26$ & $9.1688E+01$	& $4.518387773$ &	$60.04200571$ &	$0.05542$ &	$0.000488515$	\\
\hline $2.0$	 & $0.805890811$	& $1.6780E+28$ & $1.899E+26$ & $8.8362E+01$	& $4.481445362$	& $60.50853985$	& $0.05369$ &	$0.000346546$	\\
\hline $5.0$   & $0.631466547$	& $8.5910E+28$ & $1.187E+27$ & $7.2376E+01$	& $4.281871121$	& $62.34122663$	& $0.05704$ &	$0.00010214$	\\
\hline $10.0$  & $0.481933605$	& $3.7250E+29$ & $6.540E+27$ & $5.6961E+01$	& $4.042361036$	& $64.04767351$	& $0.06900$ &	$3.50461E-05$	\\
\hline	$20.0$ &	$0.342870238$ &	$1.8708E+30$ &	$4.509E+28$ &	$4.1488E+01$ & 	$3.725415157$ &	$65.9784824$ &	$0.09220$ &	$1.07177E-05$	\\
\hline	$30.0$ &	$0.273087898$ &	$5.0736E+30$ &	$1.519E+29$ &	$3.3401E+01$	& $3.508594019$ &	$67.19301334$ &	$0.113720$ &	$5.1195E-06$	\\
\hline	$40.0$	& $0.2298456$	&	$1.0484E+31$	&	$3.711E+29$	&	$2.8251E+01$	&	$3.341136905$ &	$68.08628525$	&	$0.133740$ &	$2.97431E-06$	\\
\hline $50.0$  & $0.199975$	& $1.8642E+31$ & $7.526E+29$ & $2.4771E+01$	& $3.209687665$	& $68.79328222$	& $0.153270$ &	$1.94239E-06$	\\
\hline $70.0$	&	$0.160802674$	&	$4.4862E+31$	&	$2.244E+30$	&	$1.9994E+01$	&	$2.995444763$	&	$69.88568799$	&	$0.189200$	&	$1.0013E-06$	\\
\hline $100.0$ & $0.126520749$	& $1.1600E+32$	& $7.336E+30$	& $1.5813E+01$	& $2.760814215$ &	$71.0703029$	& $0.240450$ &	$4.91113E-07$	\\

\hline     
 \end{tabular} 
 \caption{\footnotesize{Q-ball properties for $m_{s} = 100 \GeV$, $K = -0.01$ and $M_m = 10^{14} \GeV$.}}  
 \end{center}
 \end{table}

 \begin{table}[h]
\begin{center}
\begin{tabular}{|c|c|c|c|c|c|c|c|c|c|}
\hline $\varphi(0) / M_m$ & $\omega / m_s$ & $E$ (GeV) & $Q$ & $E/Q$ (GeV) & $ln(E/Q)$ & $ln(Q)$	&	$Radius$ $({\rm GeV^{-1}})$	&	$T_d$ (GeV)\\
\hline	$0.01$	&	$1.056576074$	&	$2.25989E+24$	&	$2.12484E+21$	&	$1.06356E+03$	&	$6.969374913$	&	$49.10798346$	&	$0.01629$	&	$1.492192879$	\\
\hline	$0.05$	&	$1.03614719$	&	$2.46530E+25$	&	$2.35726E+22$	&	$1.04583E+03$	&	$6.952568855$	&	$51.51437197$	&	$0.01305$	&	$0.348541733$ \\
\hline	$0.10$	&	$1.023094326$	&	$5.50176E+25$	&	$5.30691E+22$	&	$1.03672E+03$	&	$6.943813575$	&	$52.32588179$	&	$0.0111$	&	$0.193861456$	\\
\hline	$1.0$	&	$0.898900996$	&	$6.28241E+26$	&	$6.56997E+23$	&	$9.56231E+02$	&	$6.862999678$	&	$54.84196641$	&	$0.00586$	&	$0.023955186$	\\
\hline	$10.0$	&	$0.481316943$	&	$3.11960E+28$	&	$6.53960E+25$	&	$4.77032E+02$	&	$6.167584066$	&	$59.44250333$	&	$0.00582$	&	$0.000934351$	\\
\hline	$100.0$	&	$0.125039994$	&	$1.13925E+31$	&	$7.24925E+28$	&	$1.57154E+02$	&	$5.057227415$	&	$66.45328062$	&	$0.02408$	&	$1.53744E-05$ \\
 
\hline     
 \end{tabular} 
 \caption{\footnotesize{Q-ball properties for $m_{s} = 1 \TeV$, $K = -0.01$ and $M_m = 10^{14} \GeV$.}}  
 \end{center}
 \end{table}

\begin{table}[h]
\begin{center}
\begin{tabular}{|c|c|c|c|c|c|c|c|c|}
\hline $\varphi(0) / M_m$ & $\omega / m_s$ & $E$ (GeV) & $Q$ & $E/Q$ (GeV) & $ln(E/Q)$ & $ln(Q)$	&	$Radius$ $({\rm GeV^{-1}})$	&	$T_d$ (GeV)\\
\hline $0.01$  &	$1.478478948$	& $1.8930E+24$ & $1.242E+22$	& $1.5242E+02$	& $5.026610075$	& $50.87359503$	&	$0.05846$	&	$0.115940874$ \\
\hline $0.05$	 &	$1.35882302$	& $3.4749E+25$ & $2.477E+23$	& $1.4029E+02$	& $4.943687737$	& $53.86650529$	&	$0.05881$	&	$0.023011682$ \\
\hline $0.1$   &	$1.2996538$		& $1.0967E+26$ & $8.141E+23$	& $1.3471E+02$	& $4.903147927$	& $55.05637016$	&	$0.05639$	&	$0.011384673$ \\
\hline $0.25$  &	$1.20573629$	& $4.3940E+26$ & $3.466E+24$	& $1.2677E+02$	& $4.842408969$	& $56.50504342$	&	$0.04830$	&	$0.004223059$ \\
\hline $0.5$	 &	$1.115526781$	& $1.2389E+27$ & $1.038E+25$	& $1.1935E+02$	& $4.78209829$	& $57.60192311$	&	$0.04530$	&	$0.002036739$ \\
\hline $0.75$	 &	$1.051094667$	& $2.2920E+27$ & $2.018E+25$	& $1.1358E+02$	& $4.732488063$	& $58.26673425$	&	$0.04427$	&	$0.001305652$	\\
\hline $1.0$	 &	$0.999049548$	& $3.6030E+27$ & $3.299E+25$	& $1.0921E+02$  & $4.693317626$	& $58.75824672$	&	$0.04380$	&	$0.000936223$ \\
\hline $1.5$	 &	$0.915914843$	& $7.0630E+27$ & $6.933E+25$	& $1.0188E+02$  & $4.623747456$	& $59.50091994$	&	$0.04400$	&	$0.000569496$ \\
\hline $2.0$	 &	$0.850176452$	& $1.1770E+28$ & $1.225E+26$	& $9.6082E+01$	& $4.56519817$	& $60.07015326$	&	$0.04507$	&	$0.000392463$ \\
\hline $5.0$	 &	$0.61553229$	& $7.2540E+28$ & $9.974E+26$	& $7.2729E+01$	& $4.28674152$	& $62.16719412$	&	$0.05418$	&	$0.000101858$	\\
\hline $10.0$	 &	$0.437881262$	& $3.4998E+29$ & $6.584E+27$	& $5.3156E+01$	& $3.973233547$	& $64.05443997$	&	$0.07115$	&	$3.12378E-05$	\\
\hline $50.0$	 &	$0.145430396$	& $2.2275E+31$ & $1.194E+30$	& $1.8660E+01$	& $2.926407246$	& $69.25461052$	&	$0.19862$	&	$1.23958E-06$ \\
\hline $100.0$ &	$0.079536155$	& $1.5571E+32$ & $1.491E+31$	& $1.0441E+01$	& $2.345715332$	& $71.77982634$	&	$0.35589$	&	$2.54148E-07$	\\

\hline     
 \end{tabular} 
 \caption{\footnotesize{Q-ball properties for $m_{s} = 100 \GeV$, $K = -0.1$ and $M_m = 10^{14} \GeV$.}}  
 \end{center}
 \end{table}

  \section{Consequences for AD Baryogenesis and Gravitino Dark Matter in GMSB}

      There are two issues facing AD baryogenesis in GMSB models:
(i) are the Q-balls sufficiently unstable to decay before BBN and
(ii) if NLSPs are produced in Q-ball decay, can they decay to gravitinos (and, in particular, gravitino dark matter) without causing problems for BBN.

   A new possibility in GMSB is that $E/Q$ can be less than the MSSM-LSP mass but greater than the nucleon mass. This is in contrast to the case of gravity-mediated Q-balls, where $E/Q$ is approximately equal to the mass of the squarks forming the Q-ball. Since squarks generally cannot be the MSSM-LSP, in this case $E/Q > m_{\chi}$ and so the Q-balls can always decay to MSSM-LSPs. In GMSB models, on the other hand, if $E/Q < m_{\chi}$ then decay to MSSM-LSPs is kinematically excluded. Q-balls can then only decay by a process which is related to annihilation of the squarks making up the Q-ball, producing only SM fermions. In fact, some MSSM-LSPs can be produced during the latter stages of Q-ball decay, when $\varphi(0)/M_m \rightarrow 0$ and $E/Q$ increases \cite{kknew}. In this case the gravitino density will be highly suppresed relative to the $E/Q > m_{\chi}$ case, although MSSM-LSP decay could still impose BBN constraints on the gravitino mass \cite{kknew}.

    In addition, $|\Phi|/M_m$ could be significantly larger than 1 when the condensate fragments. This would allow smaller messenger masses to be consistent with unstable Q-balls. However, as we have shown, there will be an upper limit on $|\Phi|/M_m$ from the decay temperature of the Q-balls. which decreases with $\varphi(0)/M_m$ and eventually will drop below the nucleosynthesis bound. We estimate that the upper bound on $\varphi(0)/M_m$ will be at most O(100), with O(10) being more typical for realistic potential parameters.

   In general, decay of thermal relic MSSM-LSPs cannot produce gravitino dark matter, since their decay would violate BBN constraints \cite{kaz}. In addition, since the reheating temperature of the $d=6$ AD baryogenesis model is low, $T_{R} \lae 10 \GeV$ 
\cite{fd1}, gravitino dark matter cannot be generated by thermal scattering. Therefore an alternative dark matter candidate to the gravitino LSP is necessary when $E/Q < m_{\chi}$. The density of this candidate should not be diluted by the low reheating temperature. A natural possibility is the axion, whose density is effectively created at the QCD phase transition at low temperature.

    In the case where $E/Q > m_{\chi}$, GMSB Q-balls can decay to 
MSSM-LSPs. In this case, Q-ball decay to non-thermal MSSM-LSPs may be compatible with BBN. This requires that $m_{3/2} \lae 1 \GeV$ \cite{kaz}. In the case where the AD condensate evolves to purely positive charged Q-balls, the gravitino LSP mass must be $m_{3/2} \approx 2 \GeV$ in order to account for dark matter. This is slightly larger than the upper bound from BBN. If the MSSM-LSP density from Q-ball decay can be slightly enhanced relative to the baryon density,  decreasing the gravitino mass to less than 1 GeV, Q-ball decay could provide a mechanism for gravitino dark matter. This can be achieved if the AD condensate fragments to both positive and negative charged Q-balls. This has been observed in numerical simulations in the case where the original condensate fragments have more energy than Q-balls of the same charge \cite{qbf}, in which case the excess energy is converted into $\pm$Q-ball pairs.  Such large energy fragments are natural in GMSB, since the AD condensate is strongly elliptical. In this case the decay of the positive and negative charged Q-balls will increase the number of MSSM-LSPs produced per baryon number from Q-ball decay, since the MSSM-LSPs from $\pm$Q-ball decay do not cancel. The necessary enhancement of the MSSM-LSP density is very modest, a factor of 2 being sufficient to have $m_{3/2} \lae 1 \GeV$. Such an enhancement is very likely to occur, based on the numerical results of \cite{qbf}. A non-thermal MSSM-LSP density consistent with BBN could then be produced if $1 \MeV < T_{d} < T_{\chi}$. 

   Therefore $d=6$ AD baryogenesis could explain the baryon asymmetry and provide a natural source of non-thermal MSSM-LSPs and gravitino dark matter in the gauge-mediated MSSM. This can be achieved with just the MSSM, requiring no additional particles or interactions.

    To see if this Q-ball decay scenario for gravitino dark matter can be realized, we need to consider the full process of AD baryogenesis for a given set of potential parameters. We first need to compute the reheating temperature and the energy and charge of the condensate fragments. From the charge of the fragments we can estimate the charge of the resulting Q-balls and so determine the value of $E/Q$ relative to the MSSM-LSP mass and whether the Q-balls decay before nucleosynthesis but after MSSM-LSP freeze-out. We also need to check that the messenger scale is compatible with the required gravitino mass from \eq{e0}. A semi-analytical method to study 
AD baryogenesis and condensate fragmentation was given in \cite{fd1}. It is beyond the scope of the present analysis to combine this method with the Q-ball solutions discussed here, but our initial estimates indicate the above scenario for $d = 6$ AD baryogenesis can be realized with reasonable assumptions for the parameters of the potential. We will present a complete analysis of the Q-ball decay model for baryogenesis and gravitino dark matter in a future study.

\section{Conclusions}

   We have presented a new type of Q-ball solution based on the form of flat-direction potential expected in $d = 6$ AD baryogenesis in the gauge-mediated MSSM. The solution interpolates between gravity-mediated-type Q-balls with constant $E/Q$ at $\varphi(0)/M_m \ll 1$ and gauge-mediated-type Q-balls with $E/Q \propto Q^{-1/4}$ at $\varphi(0)/M_m \gg 1$. In general the gravitino mass should not be much smaller than 1 GeV in order to maintain the large messenger mass necessary for unstable Q-balls in $d=6$ AD baryogenesis. The new Q-ball solutions can be unstable for $\varphi(0)/M_m$ significantly larger than 1, with values ${\cal O}$(100) only suppressing $E/Q$ to about 15$\%$ of the AD scalar mass, which is generally much larger than the nucleon mass. Therefore taking into account the time delay for the AD condensate to fragment, the AD field can be significantly larger than the messenger scale at the onset of AD baryogenesis and still produce unstable Q-balls. This will allow smaller messenger masses and so smaller gravitino masses to be compatible with unstable Q-balls. However, there will be an upper limit on $\varphi(0)/M_m$, of order 100, from the Q-ball decay temperature and the nucleosynthesis bound.

      If $E/Q$ is less than the MSSM-LSP mass then the Q-balls decay primarily via a squark annihilation process to SM quarks, with only a small number of MSSM-LSPs being produced during the latter stages of Q-ball decay as $\varphi(0)/M_m \rightarrow 0$. In this case there is no source of gravitino dark matter, therefore an alternative candidate is necessary. This should be produced at a low enough temperature to evade dilution due to the low reheating temperature. An axion is a good candidate in this case.

       In general, gravitino dark matter in the MSSM requires either gravitinos from thermal scattering or from non-thermal MSSM-LSPs which are produced at low temperature, below the freeze-out temperature of the MSSM-LSPs. Unstable Q-balls provide a natural source of non-thermal MSSM-LSPs in the gauge-mediated MSSM with AD baryogenesis, since in the case of $d = 6$ AD baryogenesis the Q-balls typically decay at $T_{d} \lae 10 \GeV$.

      If $E/Q$ is larger than the MSSM-LSP mass, the Q-balls will decay to MSSM-LSPs. To have decaying non-thermal MSSM-LSPs which are compatible with nucleosynthesis, the gravitino must be sufficiently light, $m_{3/2} \lae 1 \GeV$ for the case of stau or sneutrino MSSM-LSPs. This requires that the AD condensate fragments to both positive and negative charged Q-balls, since in the case where only positive charged Q-balls are produced the gravitino mass is fixed by B-conservation and R-parity conservation to be $m_{3/2} \approx 2 \GeV$. In GMSB the AD condensate is elliptical, which results in positive and negative charged Q-balls in numerical simulations. Therefore the factor of two enhancement of the MSSM-LSP density necessary to have $m_{3/2} \lae 1 \GeV$ is likely to be naturally achieved.

   In this case decaying Q-balls in $d=6$ AD baryogenesis provides a remarkably minimal model for baryogenesis and gravitino dark matter in the gauge-mediated MSSM, requiring no new fields or interactions beyond the MSSM.

     We note that in the case where gravitino dark matter and the baryon asymmetry both originate from Q-ball decay, there could be some unique signatures which could distinguish the model from a generic non-thermal source of gravitino dark matter. In particular, correlated baryon and dark matter isocurvature perturbations can be naturally generated in this class of model, due to phase fluctuations of the AD field \cite{isojm,isojm2,isonew}.

    An alternative model for gravitino dark matter from Q-ball decay has been proposed in \cite{kknew}. This is based on Q-balls with $E/Q$ less than the MSSM-LSP mass, so that the main decay mode is to nucleons, with a small branching ratio to gravitinos\footnote{A model for gravitino dark matter from Q-ball decay has been proposed in \cite{kshoe}. However, this model assumes a large branching ratio to gravitinos when $E/Q$ is less than the MSSM-LSP mass, rather than the realistic small branching ratio discussed in \cite{kknew}.}. By having a sufficiently elliptical AD condensate, it is argued that a large number of$\pm$Q-balls can be produced, which, when combined with the assumed saturation of the $|\Delta Q| = 2$ annihilation mode to quarks, enhances the gravitino density relative to nucleons enough to account for gravitino dark matter. 
The Q-ball solutions we have discussed here should be relevant to the study of that model also. It will be interesting to compare the model of \cite{kknew} with ours in a complete study of the evolution of the AD condensate and its fragmentation.

   In order to fully understand how the new Q-ball solutions and their decay modes affect $d = 6$ AD baryogenesis in GMSB and the possibility of gravitino dark matter, a global analysis, which follows the evolution of the AD condensate from the beginning of its oscillations through fragmentation, Q-ball formation and decay, is necessary. We will present such an analysis in a future work.

\section*{Acknowledgements}
The work of JM is supported by the Lancaster-Manchester-Sheffield Consortium for Fundamental Physics under STFC grant
ST/J000418/1.

%%%%%%%%%%%%%%%%%%%%%%%%%%%%%%%%%%%%%%%%%%%%%%%%%%%%%%%%%%%%%%%%%%%%

\end{document}